\documentclass[11pt,a4paper]{article}
\usepackage{jcappub}
\usepackage{graphics}
\usepackage{dcolumn}
\usepackage{amsfonts}
\usepackage{amssymb}
\usepackage{bm}
\usepackage{hyperref}
\usepackage{url}
\usepackage{subfigure}
\usepackage[sort&compress]{natbib}
\usepackage{txfonts}
\usepackage{color}

\begin{document}
\newcommand{\changeR}[1]{\textcolor{red}{#1}}

\newcommand{\TBB}{{{T_{\rm BB}}}}
\newcommand{\TCMB}{{{T_{\rm CMB}}}}
\newcommand{\Te}{{{T_{\rm e}}}}
\newcommand{\Teq}{{{T^{\rm eq}_{\rm e}}}}
\newcommand{\Ti}{{{T_{\rm i}}}}
\newcommand{\nB}{{{n_{\rm B}}}}
\newcommand{\nHe}{{{n_{\rm ^4He}}}}
\newcommand{\nHet}{{{n_{\rm ^3He}}}}
\newcommand{\nHt}{{{n_{\rm { }^3H}}}}
\newcommand{\nHtw}{{{n_{\rm { }^2H}}}}
\newcommand{\nBes}{{{n_{\rm { }^7Be}}}}
\newcommand{\nLis}{{{n_{\rm { }^7Li}}}}
\newcommand{\nLisi}{{{n_{\rm { }^6Li}}}}
\newcommand{\nS}{{{n_{\rm s}}}}
\newcommand{\Teff}{{{T_{\rm eff}}}}

\newcommand{\id}{{{\rm d}}}
\newcommand{\aR}{{{a_{\rm R}}}}
\newcommand{\bR}{{{b_{\rm R}}}}
\newcommand{\neb}{{{n_{\rm eb}}}}
\newcommand{\neql}{{{n_{\rm eq}}}}
\newcommand{\kB}{{{k_{\rm B}}}}
\newcommand{\EB}{{{E_{\rm B}}}}
\newcommand{\zmin}{{{z_{\rm min}}}}
\newcommand{\zmax}{{{z_{\rm max}}}}
\newcommand{\YBEC}{{{Y_{\rm BEC}}}}
\newcommand{\YSZ}{{{Y_{\rm SZ}}}}
\newcommand{\rhob}{{{\rho_{\rm b}}}}
\newcommand{\Ne}{{{n_{\rm e}}}}
\newcommand{\sigT}{{{\sigma_{\rm T}}}}
\newcommand{\me}{{{m_{\rm e}}}}
\newcommand{\nBB}{{{n_{\rm BB}}}}

\newcommand{\kD}{{{{k_{\rm D}}}}}
\newcommand{\KC}{{{{K_{\rm C}}}}}
\newcommand{\KdC}{{{{K_{\rm dC}}}}}
\newcommand{\Kbr}{{{{K_{\rm br}}}}}
\newcommand{\zdC}{{{{z_{\rm dC}}}}}
\newcommand{\zbr}{{{{z_{\rm br}}}}}
\newcommand{\aC}{{{{a_{\rm C}}}}}
\newcommand{\adC}{{{{a_{\rm dC}}}}}
\newcommand{\abr}{{{{a_{\rm br}}}}}
\newcommand{\gdC}{{{{g_{\rm dC}}}}}
\newcommand{\gbr}{{{{g_{\rm br}}}}}
\newcommand{\gff}{{{{g_{\rm ff}}}}}
\newcommand{\xe}{{{{x_{\rm e}}}}}
\newcommand{\alphafs}{{{{\alpha_{\rm fs}}}}}
\newcommand{\YHe}{{{{Y_{\rm He}}}}}
\newcommand{\SE}{{{\dot{{\mathcal{E}}}}}}
\newcommand{\SQ}{{{{{\mathcal{E}}}}}}
\newcommand{\SN}{{\dot{\mathcal{N}}}}
\newcommand{\Sn}{{{\mathcal{N}}}}
\newcommand{\muc}{{{{\mu_{\rm c}}}}}
\newcommand{\xc}{{{{x_{\rm c}}}}}
\newcommand{\xH}{{{{x_{\rm H}}}}}
\newcommand{\mT}{{{{\mathcal{T}}}}}
\newcommand{\Ob}{{{{\Omega_{\rm b}}}}}
\newcommand{\Or}{{{{\Omega_{\rm r}}}}}
\newcommand{\Odm}{{{{\Omega_{\rm dm}}}}}
\newcommand{\mdm}{{{{m_{\rm WIMP}}}}}

\title{Creation of the CMB spectrum: precise analytic solutions for
  the blackbody photosphere}

\author[a]{Rishi Khatri,}
\author[a,b,c]{Rashid A. Sunyaev}

\affiliation[a]{ Max Planck Institut f\"{u}r Astrophysik\\, Karl-Schwarzschild-Str. 1
  85741, Garching, Germany }
\affiliation[b]{Space Research Institute, Russian Academy of Sciences, Profsoyuznaya
 84/32, 117997 Moscow, Russia}
\affiliation[c]{Institute for Advanced Study, Einstein Drive, Princeton, New Jersey 08540, USA}
\date{\today}
\emailAdd{khatri@mpa-garching.mpg.de}
\abstract
{
The blackbody spectrum of CMB was created in the blackbody photosphere at redshifts  $z\gtrsim
2\times 10^6$.  At these early times, the Universe was dense and hot enough
that complete thermal equilibrium between baryonic matter (electrons and
ions) and photons could be established on time scales much shorter than the age
of the Universe.  Any perturbation away from the
blackbody spectrum was  
 suppressed exponentially. New physics, for example annihilation and decay of dark matter,
can add energy and photons to CMB at redshifts $z\gtrsim  10^5$ and result in a Bose-Einstein spectrum
with a non-zero chemical potential ($\mu$).  Precise evolution of the
CMB spectrum  around the critical redshift of $z\simeq 2\times 10^6$ is
required in order to calculate the $\mu$-type spectral distortion and
constrain the underlying new physics.
Although numerical calculation of important processes involved (double
Compton process, comptonization and bremsstrahlung) is not difficult
with present day computers,
analytic solutions are much faster and easier  to calculate and  provide valuable
physical insights. We provide precise (better than $1\%$) analytic
solutions for the decay of $\mu$, created at an earlier epoch, including all three processes, double Compton, Compton scattering
on thermal electrons and
bremsstrahlung in the limit of small distortions. This is a significant improvement over the existing  solutions with
accuracy   $\sim 10\%$ or worse.  We  also give a census of important sources of
energy injection into CMB in standard cosmology. In particular,
  calculations of distortions from electron-positron annihilation and primordial nucleosynthesis  
illustrate in a dramatic way 
the strength of the equilibrium restoring processes in the early
Universe. Finally, we point out the triple degeneracy in standard
cosmology, i.e., the $\mu$ and $y$ distortions from adiabatic cooling of
baryons and electrons,
Silk damping and annihilation of  thermally produced WIMP dark
matter are of similar order of magnitude ($\sim 10^{-8}-10^{-10}$).
}

\maketitle
\flushbottom
\section{Introduction}
There are several important events in the history of the Universe which
provide the foundations for the standard cosmological model. The first
event is the \emph{big bang}, which created the present expanding Universe
filled with matter, radiation and dark energy.    One of the goals of cosmology is to reconstruct these
initial conditions and learn about the high energy physics at early times.
We should emphasize here that the blackbody spectrum of cosmic microwave
background (CMB) is \emph{not} an initial condition. The blackbody spectrum
is created and maintained dynamically throughout the early history of the
Universe ($z\gtrsim 2\times 10^6$) by standard physics processes and how
this happens is the main topic of the present paper. The CMB 
spectrum thus provides information about the physics of the Universe after
the big bang. 

\begin{figure*}
\includegraphics[width=15cm]{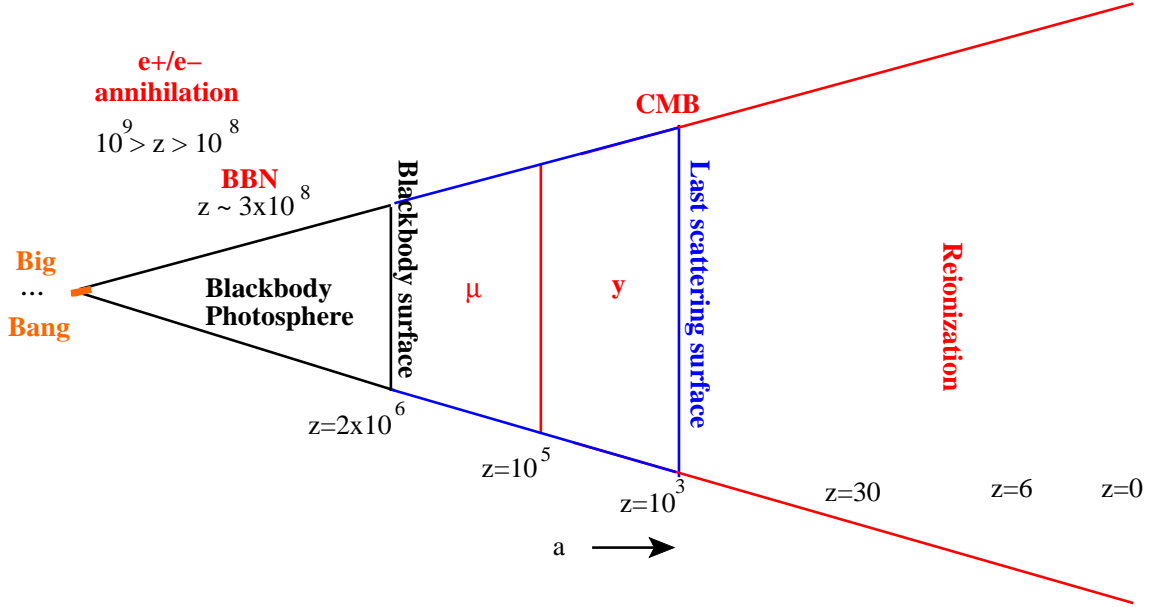}
\caption{\label{sketch}{Important events in the history of the CMB spectrum
  and anisotropy formation  in
  big bang cosmology. Redshift range ($2\times 10^6\gtrsim
    z\gtrsim 10^5$), where the energy injection would give rise to a
    Bose-Einstein spectrum ($\mu$-type distortion),  is marked as $\mu$. At much smaller redshifts ($z\lesssim 10^4$), any heating of CMB
  through Compton scattering would  create  a $y$-type distortion. The
  spectrum in the intermediate redshift range would not be a pure $\mu$ or
  $y$ type but in between the two types.}}
\end{figure*}

The other important events in standard cosmology are: the formation of light elements in primordial
nucleosynthesis \citep{wfh67}, recombination of electrons and ions to neutral
atoms \citep{zks68,peebles68}, and reionization of
the Universe \citep{gp1965} by the radiation emitted by  first stars and galaxies. We should also mention
electron-positron annihilation at $10^{10} \gtrsim z\gtrsim 10^8$  which more than doubles the
entropy of CMB and raises its temperature by $\sim 40\%$. The important
events in the history of the Universe are sketched in Fig. \ref{sketch}.
Big bang nucleosynthesis (BBN) theory together with measurement of light
elements abundances constrains the photon to baryon number density at
$z\sim 10^8$ \citep[see][for a review]{iocco2009} and is the earliest direct evidence and
measurement 
 of electromagnetic radiation in the early Universe. The fact that   the
 photon energy density inferred from BBN and Cosmic microwave background
 (CMB) at $z\sim 10^3$ \citep{wmap7} are close to each other (consistent within
 $\sim 2-\sigma$)
 also implies that we do not have arbitrary freedom in adding energy to CMB
 between these two epochs. {There is, however, no direct way to constrain
 the energy density in photons before  BBN and the epoch of electron-positron
 annihilation.}

{The CMB (anisotropy and polarization)} is at present the most precise cosmological
probe. The CMB spectrum was created at $z\gtrsim 2\times 10^{6}$, and this critical
redshift defines the \emph{blackbody surface} for our Universe. Spatial fluctuations in the
temperature of the CMB were imprinted much later, when the electron and protons
recombined to form hydrogen atoms at $z\approx 1100$. This second boundary
defines the well known \emph{last scattering surface}, the structure of
which is encoded in   the photon visibility function \citep[first studied
by][]{sz1970c}. The anisotropies and
physics at the last scattering surface have been very well studied and
accurate analytic \citep{hu1995} and numerical solutions
\citep[]{mabert95,cmbfast} have been available for some time, motivated by
the precise experiments such as WMAP \citep{wmap7} and Planck
\citep{planck}. 

{The blackbody spectrum, once created at high redshifts (for
  example before the time of electron-positron annihilation), is preserved by the
adiabatic expansion of the Universe at all subsequent times. However, if there is energy release at
lower redshifts , for example by particle decay and annihilation or Silk
damping, it will distort the CMB spectrum away from the Planck form.
In this case, Zeldovich and Sunyaev \citep{zs1969} first demonstrated that bremsstrahlung alone cannot
recreate  blackbody spectrum until very high redshifts, almost up to  the time of
electron-positron annihilation. The problem
of evolution of the CMB spectrum through the blackbody surface, in the
presence of heating,  was solved
analytically by \citep{sz1970} including the processes of comptonization
 and emission and absorption of photons with special emphasis on
 bremsstrahlung.  Comptonization  is the process of redistribution of
 photons over frequency, resulting from the Doppler and recoil effects of
 Compton scattering of photons on
thermal electrons \citep{k1956}.  Since the double Compton cross section
\citep{lightman,thorne81} has a dependence on
frequency similar to bremsstrahlung, the solution of Sunyaev and Zeldovich \citep{sz1970}   also allowed 
inclusion of  double
Compton emission and absorption, {which is dominant over
  bremsstrahlung in a low baryon density Universe such as ours and was
  first considered by \citep{dd1982}.} Double Compton emission or
 absorption is just the first radiative correction to the process of
 Compton scattering just as bremsstrahlung is the first radiative
 correction to the scattering of electrons on nuclei \citep[see discussion
 in ][]{pss1983}.} The approximations used in these
analytic solutions result in accuracy of $\sim 5-10\%$ at  redshifts
($z\lesssim 3\times 10^6$) and
worse than $10\%$ at higher redshifts. Much better accuracy
is, of course, achievable in numerical solutions \citep{is1975,bdd91,hs1993,pb2009,cs2011}.

The type of spectrum we get for  the CMB 
   is
determined by the Compton $y$ parameter, defined in Eq. \eqref{yz}. Bose-Einstein spectrum can be created only at high redshifts
$z\gtrsim 10^5, y>1$ \citep{sz1970c}, when comptonization of CMB is very efficient. Any addition of energy and photons to
CMB at $z\lesssim  10^5, y<<1$    inevitably distorts the CMB
spectrum with a $y$-type distortion \citep{zs1969}. The $y$-type distortions, in contrast to $\mu$-type distortions,  can be created
at low redshifts up to $z=0$.  Reionization at $z\sim 10$ in particular is expected to
create $y$-type distortions of magnitude $y_{\rm e}\sim \tau_{ri}\kB\Te/(\me
c^2)\sim 10^{-7}$ where $\tau_{ri}\approx 0.1$ is the optical depth due to
reionization, $\Te\approx 10^4$ is the average temperature during
reionization, $\kB$ is the Boltzmann's constant, $c$ is the speed of light
and $\me$ is the mass of electron.  Due to uncertainties in reionization physics, it will be
extremely difficult to separate these low redshift distortions from the
$y$-type distortions created before recombination. COBE FIRAS experiment \citep{cobe} measured the CMB spectrum with high
precision and placed a constraint on the chemical potential of
Bose-Einstein spectrum for CMB of $\mu\lesssim 9\times 10^{-5}$, thus
confirming that the CMB has a Planck spectrum at very high
accuracy.

In standard model of cosmology we can get constraints on the energy density of
radiation from two distinct and very precise observables. The first is the
deuterium abundance, which gives the baryon number  to photon number
ratio $\eta=(5.7\pm 0.3)\times 10^{-10}$ during
primordial nucleosynthesis at $4\times 10^8 \gtrsim z\gtrsim 4\times 10^7$ \citep{fields2001,iocco2009}.  The second is the
measurement of CMB anisotropies, which constrains the baryon to photon
ratio $\eta=(6.18\pm 0.15)\times  10^{-10}$ \citep{wmap7} during recombination at $z\approx 1100$. The fact that these
two values of baryon to photon ratio are almost identical with  small
error bars means that we do not have arbitrary freedom in adding energy to
CMB, for example, with the
introduction of new physics. {In fact any addition of energy/entropy to
CMB between primordial nucleosynthesis and recombination cannot be more
than a small percentage ($\lesssim 10\%$) of the photon energy density during BBN.}

In standard
cosmology, the chemical potential of CMB is expected to have a magnitude of
$\mu\sim 10^{-9}-10^{-8}$ resulting from  the heating of CMB by Silk damping.
A similar magnitude but opposite sign distortion, $\mu\approx -2.7\times 10^9$, is expected from  cooling
of photons due to energy losses to baryons and electrons, which have a
 different adiabatic index (5/3) than radiation (4/3) and as a result cool
 faster than the radiation as the Universe expands \citep{cs2011,ksc2011}. 
A detection of a chemical potential
of magnitude greater than $10^{-8}$ will therefore mean existence of
non-standard physics {(or a small scale power spectrum that is bigger than what is
expected from extrapolating the large scale power spectrum as measured by
CMB and large scale structure)} at these redshifts. Proposed experiment Pixie
\citep{pixie} has exactly this level of sensitivity. Constraining high
energy physics using $\mu$-type distortions requires precise calculation of
evolution of the CMB spectrum through the blackbody surface at $z\approx 2\times
10^6$. Analytic solutions provide valuable physical insight in addition to
being much easier to compute. Motivated by these factors we try to improve the
method of \citep{sz1970} to  achieve better than $1\%$ precision in
analytic solutions. These solutions should prove useful in predicting the
signal from models of high energy physics which can provide  energy
injection to CMB at these high redshifts.

In the last part of the paper we  apply our analytic solution to examples from
the standard model of cosmology including electron-positron annihilation and
dark matter annihilation. These simple examples in particular demonstrate
the efficiency of processes responsible for maintaining thermal equilibrium in
the early Universe. They also demonstrate the difficulty of creating
spectral distortion in CMB at high redshifts. There remains thus only a narrow
window at redshifts $10^5<z<2\times 10^6$ when a Bose-Einstein spectrum can
be created.

We  use WMAP \citep{wmap7} best fit parameters for ${\rm \Lambda CDM}$
cosmology for numerical calculations.

\section{Thermalization of CMB}
Compton scattering  is responsible for creating a
Bose-Einstein spectrum of photons if the rate of comptonization {(i.e. the
redistribution of photons over the entire spectrum by Compton scattering)} is greater than
the expansion (Hubble) rate of the Universe \citep{zs1969}. This condition is satisfied at
redshifts $z\gtrsim 10^5$. In the non-relativistic regime
comptonization is described by Kompaneets equation \citep{k1956}. Compton
scattering, however, conserves photon number and therefore the spectrum
obtained as a result of comptonization will in general have a non-zero
chemical potential. At redshifts $z\gtrsim 2\times 10^6$ double Compton
scattering, and to a lesser extent bremsstrahlung, can emit/absorb photons
at low frequencies very efficiently because  {the
  optical depth and the absorption coefficient} of these two processes increase with decreasing frequency as $\propto \nu^{-2}$. Compton scattering is then able to
redistribute these photons over the entire spectrum. The net result is that
the chemical potential is exponentially suppressed and a Planck spectrum is
established.

\begin{figure}
\includegraphics[width=14cm]{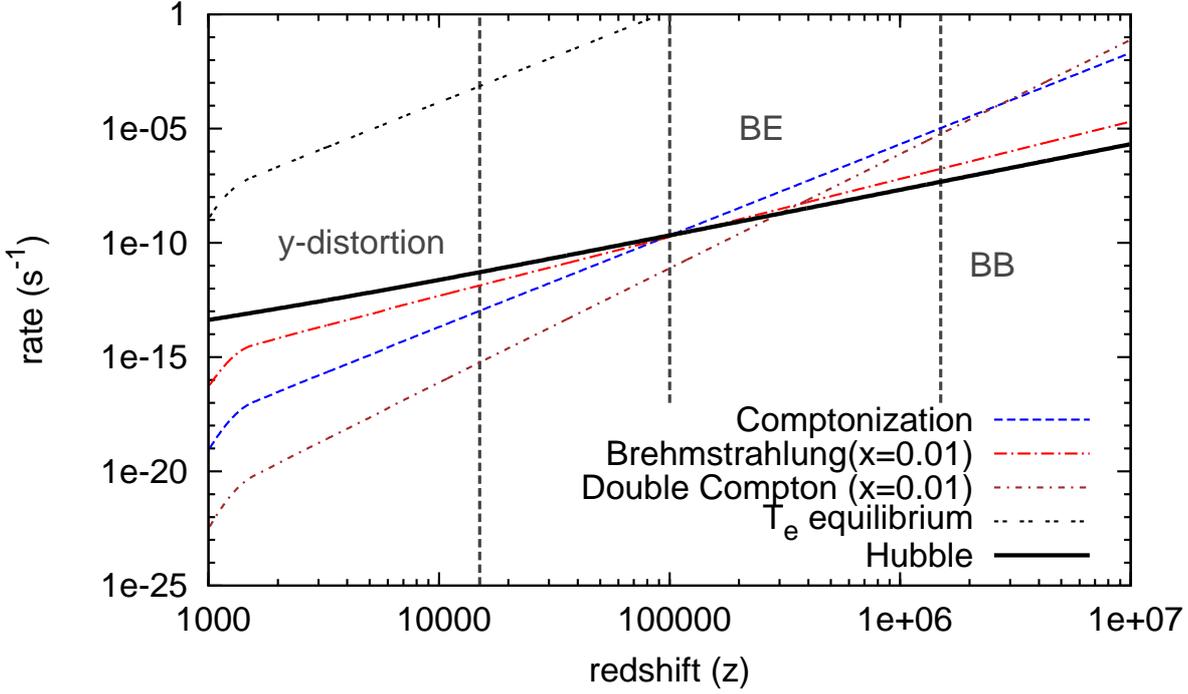}
\caption{\label{rates}Rates of bremsstrahlung, comptonization and double Compton
  scattering. Bremsstrahlung and double Compton absorption have a dependence
  on frequency $\sim  x^{-2}$. Rates above are for $\xe=0.01$ which is the
  typical value of critical frequency at which double Compton/
  bremsstrahlung absorption becomes equal to comptonization rate
  (Eq. \eqref{xc} and Fig. \ref{xrates}). Also shown is
  the rate at which electron/baryon plasma achieves equilibrium temperature
$\Te$ given by Eq. \eqref{te}.  $\Te=T_{\gamma}$ and blackbody spectrum is
assumed for this figure. Compton scattering is able to create Bose-Einstein
spectrum at $z\gtrsim 10^5$ while  for redshifts $z\lesssim 15000$ any energy injection
gives rise to a $y$-type distortion \citep{zs1969}. The distortion has a
shape in-between $y$-type and $\mu$-type for redshifts $1.5\times 10^4\lesssim
z\lesssim 10^5$ and we choose $z=5\times 10^4$ to divide our estimates
between $\mu$-type and $y$-type.}
\end{figure}

The equilibrium electron temperature in a radiation field with occupation
number $n(x)$, where $x\equiv h\nu/kT_{\gamma}$ and $T_{\gamma}$ is the
temperature of reference blackbody, is given by \citep[]{zl1970,ls1971}
\begin{align}
\frac{\Te}{T_{\gamma}}=\frac{\int(n+n^2)x^4\id x}{4\int n x^3 \id x}\label{te}
\end{align}
The
rate ($8 \sigT E_{\gamma}/(3 \me c)\times  \Ne/(\Ne+n_i)$) at which electron/baryon plasma achieves equilibrium temperature
$\Te$ given by Eq. \eqref{te} is shown
in Fig. \ref{rates} in the topmost curve. 
Rates of bremsstrahlung absorption $\Kbr(e^{\xe}-1)/\xe^3$, double
Compton absorption $\KdC(e^{\xe}-1)/\xe^3$ and
Compton scattering $\KC$ 
are also compared with the Hubble rate in Fig. \ref{rates}. $\Ne$ is the
electron number density, $\sigT$ is the Thomson cross section,   $\Te$ is
the electron temperature,   $n_i$ is the number density of ions,  $\xe=h\nu/\kB \Te$ is the dimensionless frequency
corresponding to frequency $\nu$, $h$ is the Planck's constant,
$E_{\gamma}$ is the energy density of photons. {$\Te=T_{\gamma}=\TCMB(1+z)$,
$\TCMB=2.725{\rm K}$ and  blackbody spectrum is
assumed for this figure.} The rate coefficients  $\KC,\KdC,\Kbr$ are given by:
\begin{align}
\KC&=\Ne\sigT c\frac{\kB \Te}{\me c^2}\nonumber\\
&\equiv  \aC(1+z)^4=2.045\times 10^{-30}(1+z)^4\left(\frac{\Ob h_0^2}{0.0226}\right)\left(\frac{1-Y_{He}/2}{0.88}\right){\rm s}^{-1}\\
\KdC&=\Ne\sigT c \frac{4\alphafs}{3\pi} \left(\frac{\kB \Te}{\me
    c^2}\right)^2 \gdC(\xe)I_{dC}\nonumber\\
&\equiv  \adC(1+z)^5=7.561\times 10^{-41}(1+z)^5\gdC(\xe)
\left(\frac{\Ob
    h_0^2}{0.0226}\right)\left(\frac{1-Y_{He}/2}{0.88}\right){\rm s}^{-1}\\
\Kbr&=\Ne\sigT c\frac{\alphafs \nB}{(24\pi^3)^{1/2}}\left(\frac{\kB
\Te}{\me c^2}\right)^{-7/2}\left(\frac{h}{\me
c}\right)^3\gbr(\xe,\Te)\nonumber\\
&\equiv  \abr(1+z)^{5/2}=2.074\times 10^{-27}(1+z)^{5/2}\gbr(\xe,\Te)\left(\frac{\Ob h_0^2}{0.0226}\right)^2\left(\frac{1-Y_{He}/2}{0.88}\right){\rm s^{-1}}.
\end{align}
$\alphafs$ is the fine structure constant and $\nB$ is the baryon number
density. $I_{dC}=\int \id{\xe}~ \xe^4n(1+n)\approx 25.976$ for a blackbody
spectrum at temperature $\Te$, $\Ob$ is the baryon density parameter,
$h_0$ is the Hubble parameter and $\YHe$ is the primordial helium mass fraction. $\gbr=\sum_i Z_i^2n_i\gff(Z_i,\xe,\Te)/\nB$ is the average
gaunt factor for bremsstrahlung, $n_i$ is the number density of ion species
$i$ and $Z_i$ the charge of ion. The sum is over all ionic species, which
for the primordial plasma at high redshifts consists of protons and helium
nuclei. Accurate fitting formulas for $\gff$ have been provided by
\citep{itoh2000}. $\gdC$ is the gaunt factor for double Compton scattering and
accurate fitting formula for it has been calculated recently by
\citep{cs2011}. We  use these fitting formulae for numerical solution.
For reference at $\xe=0.01, z=2\times 10^6$, $\gdC=1.005$ and
$\gbr=2.99$ and we use these values for our analytic solutions.
These gaunt factors are slowly varying functions of time and frequency and
can be assumed to be constant in the redshift range of interest for
analytic calculations. We 
justify this assumption below. 

 The division into $y$-type and Bose-Einstein  regions depends on
the Compton $y$ parameter, 
\begin{align}
y(z)=\int_{0}^{z}\id z\frac{\kB \sigT}{\me
  c}\frac{\Ne T_{\gamma}}{H(1+z)}{.}\label{yz}
\end{align}
During radiation domination ($z\gg z_{\rm eq}=3.2\times 10^4$) the integral
  can be carried out  analytically, giving
$y(z)\approx 4.9\times 10^{-11}(1+z)^2$, where $z_{\rm eq}$ is the redshift
when matter energy density equals radiation energy density and we have
assumed  $3.046$ effective
species of massless neutrinos \citep{mm2005}.
We have, for WMAP  $\Lambda CDM$ cosmological parameters \citep{wmap7}, $y(1.5\times 10^4)\approx 0.01$, $y(4.7\times 10^4)\approx 0.1$ and
$y(1.5\times 10^5)\approx 1$. For $y\lesssim 0.01<<1$ we have a $y$-type distortion and for
$y\gtrsim 1$ a Bose-Einstein spectrum. The spectrum attained for
$0.01\lesssim y\lesssim 1$ is in between a pure $y$-type and
Bose-Einstein. We choose $z=5\times 10^4$ as an approximate division
between the two types of distortions for our estimates.
This division is accurate if the energy injection between $1.5\times
10^4\lesssim z\lesssim 1.5\times 10^5$ is small compared to the total energy injection at
$z\lesssim 2\times 10^6$.  If most of the energy injection happens  between $1.5\times
10^4\lesssim z\lesssim 1.5\times 10^5$, then numerical calculations must be
performed to get accurate  final spectrum. We note that for small distortions, we can
calculate distortions arising from different physical processes separately
and add them linearly.

\subsection{Kinetic equation}
We will follow \citep{sz1970} in solving the kinetic equation for photon
distribution using stationarity assumption and then do an iteration to
relax this assumption to arrive at a more accurate  solution.
The kinetic equation for the evolution of photon occupation number $n(\xe)$ in
the presence of Compton scattering, double Compton scattering and
bremsstrahlung is given by \citep[see also][]{bdd91,hs1993}
\begin{align}
\frac{\partial n(\xe,t)}{\partial t}&=\KC\frac{1}{\xe^2}\frac{\partial
  }{\partial \xe}\xe^4\left[\frac{\partial n}{\partial
      \xe}+n+n^2\right]+\left(\KdC+\Kbr\right)\frac{e^{-\xe}}{\xe^3}\left[1-n(e^{\xe}-1)\right]+\xe\frac{\partial n}{\partial \xe}\frac{\partial} {\partial t}\left[\ln \left(\frac{\Te}{\TCMB (1+z)}\right)\right]\label{kineticeq}
\end{align}
The first term with coefficient $\KC$
is the Kompaneets term describing Compton scattering. The three terms in
the square brackets in Kompaneets term describe photon diffusion in
frequency due to the Doppler effect, electron recoil
and induced recoil effects respectively. The second term represents
emission and absorption of photons due to double Compton ($\KdC$) and
bremsstrahlung ($\Kbr$). 
The last term in Eq. \eqref{kineticeq}  arises because we are evaluating
the time derivative at constant $\xe=h \nu/\kB\Te$ instead of constant
frequency $\nu$, and electron temperature changes with time\footnote{This
  is in addition to
the usual $(1+z)$ dependence due to the expansion of the Universe. The
variable $\xe$ is invariant w.r.t. the expansion of the Universe.} because
the photon distribution is evolving (see Eq. \eqref{te}).

We are interested in the regime where comptonization is efficient and
deviation from a Planck spectrum is small. In
this regime 
the photon spectrum  is  described  by a Bose-Einstein distribution with a
chemical potential much smaller than unity in magnitude.
\begin{align}
n(\xe)&=\frac{1}{e^{\xe+\mu}-1}\approx \frac{1}{e^{\xe}-1}-\mu\frac{e^{\xe}}{(e^{\xe}-1)^2}
\end{align}
The total energy density and number density of photons is then given by
\footnote{{The chemical potential is not constant but a function of frequency at low
    frequencies but this dependence is important only for calculating the
    photon production rate.  We can ignore the frequency dependence in
    calculating the total energy and number density in the spectrum since
    the 
    contribution from low frequencies ($x\ll 1$) to these quantities is
    small and the constant $\mu$ assumption introduces negligible error.}}
\begin{align}
E&=\frac{\aR\Te^4}{I_3}\int \id \xe \xe^3 n(\xe)\approx\aR\Te^4\left(1-\mu\frac{6\zeta (3)}{I_3}\right)\label{energy}\\
N&=\frac{\bR\Te^3}{I_2}\int \id \xe \xe^2
n(\xe)\approx\bR\Te^3\left(1-\mu\frac{\pi^2}{3I_2}\right),\label{number}
\end{align}
where $a_R=\frac{8 \pi^5\kB^4}{15 c^3 h^3}$ is the radiation constant,
$\bR=\frac{16 \pi\kB^3\zeta(3)}{ c^3 h^3}$, $\zeta$ is the Riemann zeta
function with $\zeta(3)\approx 1.20206$,  $I_3=\int x^3 n_{\rm pl}(x) \id x=\pi^4/15$,$I_2=\int x^2 n_{\rm
  pl}(x) \id x=2\zeta(3)$, and $n_{\rm pl}(x)=1/(e^x-1)$.

{In order to cancel the effect of the expansion of the Universe we will
  use the  
  blackbody spectrum with temperature $T_{\gamma}=\TCMB(1+z)$, $E_{\gamma}$
  is its energy density and $N_{\gamma}$ its
  number density. 
If we have a source injecting energy density at a rate $\SE$ and photon
number density  at rate
$\SN$, where $\SQ=E/E_{\gamma}$, $\Sn=N/N_{\gamma}$,}
we get using Eqs. \eqref{energy} and \eqref{number} and with $|\mu|<<1$
\begin{align}
\frac{\id }{\id t}\ln\left(\frac{E}{E_{\gamma}}\right)&\equiv\frac{\SE}{\SQ}=4\frac{\id  }{\id t}\ln\left(\frac{\Te}{T_{\gamma}}\right)-\frac{6\zeta
  (3)}{I_3}\frac{\id \mu}{\id t}\label{edt}\\
\frac{\id }{\id t}\ln\left(\frac{N}{N_{\gamma}}\right)&\equiv\frac{\SN}{\Sn}=3\frac{\id }{\id t}\ln\left(\frac{\Te}{T_{\gamma}}\right)-\frac{\pi^2}{3I_2}\frac{\id \mu}{\id t}\label{ndt}.
\end{align}
{ For small distortions $E\approx E_{\gamma}, N\approx N_{\gamma}$
  and we have at lowest order, $\SE/\SQ\approx \SE$ and $\SN/\Sn\approx \SN$.
The photon production due to double Compton and bremsstrahlung can be
calculated by taking the time derivative of Eq. \eqref{number}. Multiplying  the kinetic equation Eq. \eqref{kineticeq} by
$\xe^2$ and integrating over $\xe$ and using it in the the time derivative
of Eq. \eqref{number}, the terms involving $\partial \Te/\partial t$ cancel
out and only the bremsstrahlung and double
Compton terms contribute, giving}
\begin{align}
\frac{\id }{\id t}\ln\left(\frac{N}{N_{\gamma}}\right)&=\frac{1}{I_2}\int_0^{\infty}\id \xe
\left(\KdC+\Kbr\right)\frac{e^{-\xe}}{\xe}\left[1-n(e^{\xe}-1)\right]\nonumber\\
&\approx\frac{\left(\KdC+\Kbr\right)}{I_2}\int\id \xe\frac{\mu}{\xe\left(e^{\xe}-1\right)}.
\label{ndtint}
\end{align}
In general there may be additional sources of photons, for example, if energy
is injected as an electromagnetic shower resulting from decay of a heavy
particle, the resulting cascade may produce non-negligible amount of
photons. However in most cases of interest and at high redshifts this is  much smaller than the
photon production from bremsstrahlung and double Compton.

\subsection{Numerical solution}
 In order to  calculate the precision of the analytic formulae derived
  below, we compare them with the numerical solution of the coupled system of Eqs. \eqref{kineticeq} and \eqref{te}. The 
   initial spectrum for the numerical solution is a $\mu$ type distortion of magnitude $10^{-5}$ at
   $z=5\times 10^6$ 
  with the chemical potential decaying exponentially with decreasing
  frequencies at $x<<1$ (see Eq. \eqref{xc} below). The  results are not sensitive to the exact form
  of initial spectrum if the starting redshift is $\gg 10^6$, what matters
  is the total energy input into an initial blackbody. We solve equation
  \eqref{kineticeq} iteratively in small time steps (using Compton $y$
  parameter as the time variable) of $\delta y = 0.1$. In
  the first iteration we use the analytic solution for the evolution of
  electron temperature. In the second iteration we use the solution of
  first iteration to calculate the electron temperature using Eq. \eqref{te}. As a frequency variable we use a
  $x=h\nu/\kB T$ where $T$ is the reference  temperature which evolves just
  by redshifting due to cosmological expansion and is equal to the electron
  temperature at the start of each iteration step. This gets rid of the last
  term in Eq. \eqref{kineticeq} and introduces  factors of $\Te/T$ in the
  Kompaneets and bremsstrahlung/double Compton terms. We also write the
  total spectrum as a sum of blackbody part at reference temperature $T$
  and a distortion part, keep only the terms linear in distortions and
  solve the linearized Eq. \eqref{kineticeq} for distortions, as the zeroth
  order blackbody part vanishes. 
Implicit backward 
   differentiation method is used to solve the PDE. We use logarithm of $x$ as the
   second independent variable with the variable step size in the $x$
   direction. 

The main source of error is the deviation of electron
   temperature used in solving the PDE from the correct temperature given
   by Eq. \eqref{te},
   resulting in violation of energy conservation. In our solution, the maximum
  error in energy conservation as a fraction of energy in the distortion
  occurs at
  high redshifts, when there is strong evolution of $\mu$, but even this is
   $\sim 10^{-5}$ in each iterative step. The  error with respect to
   the total energy density in photons is, therefore, $\sim 10^{-10}$, since we start
   with an initial distortion of $\sim 10^{-5}$, and is much smaller in the
   later steps as the distortion decays exponentially. An important point
  to note here is that since we change the reference temperature to the current
  electron temperature at the beginning of each iterative step, and solve
  and track
   only the distortion part, the error in $\mu$ in individual iterative steps
  does not accumulate but is in fact suppressed in the subsequent evolution
  of the spectrum by the visibility factor (defined below). The solution obtained can thus be considered almost exact for the
  purpose of the present paper.

  The high frequency
  spectrum is forced to be Wien at $120\ge x>100$ with the chemical potential and
  temperature given by analytic solution. The low frequency
  boundary is at $x=10^{-5}$ and the spectrum at $x<2\times 10^{-5}$ is
  forced to be blackbody with the temperature equal to the analytic
  electron temperature. Since our boundaries are far away in the distant
  Wien/Rayleigh-Jeans tails, where there are negligible amount of photons/energy
  , the solution is not sensitive to the exact boundary
  conditions.  Since our
  boundary conditions are approximately equal to the true solution, we are
  able to use large steps in the $x$ direction resulting in considerable
  speedup in the numerical calculation compared to a calculation with
  arbitrary (but reasonable) boundary conditions.  The initial spectrum is
  evolved until recombination ($z=1100$) although the $\mu$ distortion at $x\gtrsim
  0.1$ is effectively frozen-in after $z\sim 10^4$. Further details on
  numerical issues can be found in \cite{bdd91,pb2009,cs2011}.

\section{Solution using stationarity approximation}
\begin{figure}
\includegraphics[width=14cm]{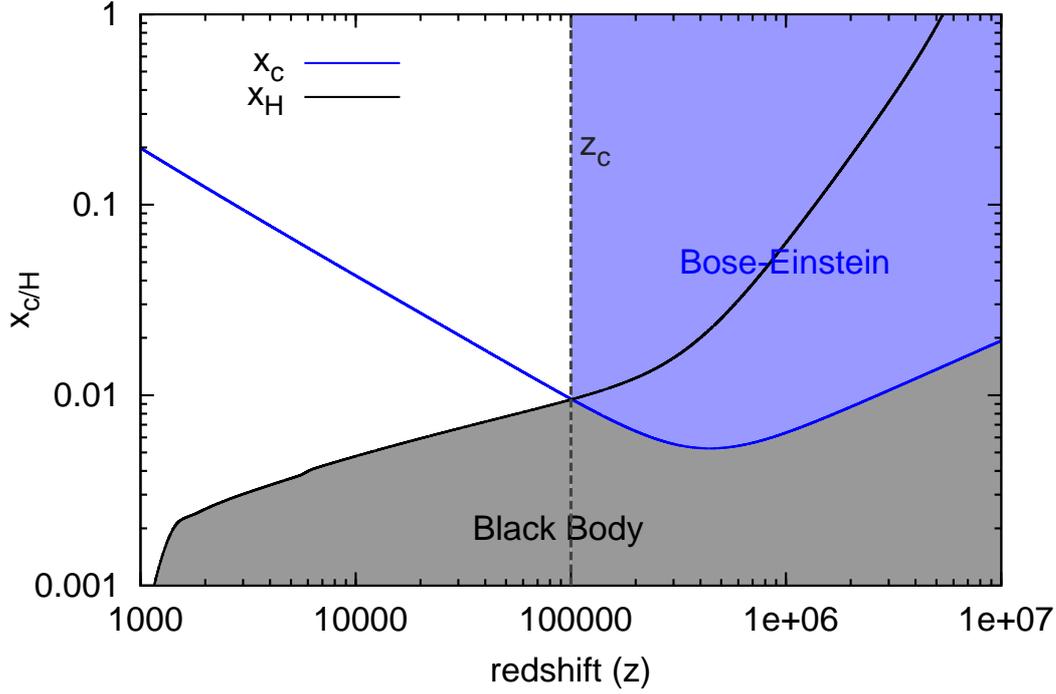}
\caption{\label{xrates}The frequency $\xc$ and $\xH$ at which the
  Compton and Hubble  rates respectively are equal
to the sum  of the bremsstrahlung and double Compton rate. If Compton
rate is also greater than the Hubble rate, then bremsstrahlung and double
Compton can establish 
complete thermodynamic equilibrium (blackbody spectrum) below $\xc$
otherwise complete thermodynamic equilibrium is established below
$\xH$. {Above $\xc$, at redshifts $z>z_{\rm c}\approx 10^5$,
 a frequency dependent 
chemical potential $\mu$ is established. At $x>>x_c$, the chemical potential has an almost
constant (frequency independent)
value $\muc$  and we have thus a Bose-Einstein spectrum. The chemical
potential decreases with time due to 
 the photons created by bremsstrahlung
and double Compton at low frequencies and redistributed by Compton
scattering over the entire spectrum.}}
\end{figure}

In order to calculate the integral in Eq. \eqref{ndtint} we need to know the
occupation number $n(\xe)$. Note that we cannot use Bose-Einstein
distribution as an approximation since the integral in this case diverges
at small $\xe$. The reason is that double Compton and bremsstrahlung rates
diverge at small frequencies and establish Planck spectrum. We can get an
approximate solution from the kinetic equation Eq. \eqref{kineticeq} \citep{k1956} by assuming that the instantaneous
spectrum is stationary. Thus, neglecting time derivatives, making chemical
potential function of frequency,\footnote{We note that any
  spectrum can be described by a frequency dependent chemical potential.} $\mu(\xe)$, and also assuming,
$\xe<<1,\mu(\xe)<<1$ Eq. \eqref{kineticeq} simplifies considerably,

\begin{align}
0&=-\KC\frac{1}{\xe^2}\frac{\id
  }{\id \xe}\xe^2\frac{\id \mu}{\id
      \xe}+\left(\KdC+\Kbr\right)\frac{\mu}{\xe^4}\label{stateq}
\end{align}
The solution of this ordinary differential equation with the boundary
condition $\mu(0)=0$ is given by \citep{sz1970}
\begin{align}
\mu(\xe)&=\muc e^{-\xc/\xe}\nonumber\\
\xc&\approx \left(\frac{\KdC(\xc)+\Kbr(\xc)}{\KC}\right)^{1/2}\nonumber\\
&=\left(\frac{\adC(1+z)+\abr(1+z)^{-3/2}}{\aC}\right)^{1/2}\nonumber\\
&\approx
\left(7.43\times 10^{-5}\left(\frac{1+z}{2\times 10^6}\right)+1.07\times
  10^{-6}\left(\frac{1+z}{2\times 10^6}\right)^{-3/2}\right)^{1/2},\label{xc}
\end{align}
where $\muc$ is normalization specified by chemical potential at large
$\xe$ .
Thus $\mu(\xe)$ decays exponentially at small frequencies and goes to
constant at large frequencies.
$\xc$ is also the frequency at which comptonization rate is equal to photon
absorption rate due to double Compton and bremsstrahlung. Similarly we can
also define a frequency at which the photon absorption rate is equal to
the Hubble rate $H(z)$ {\citep{hs1993},}
\begin{align}
\xH&\approx \left(\frac{\KdC(\xH)+\Kbr(\xH)}{H}\right)^{1/2}\nonumber\\
&= \left(\frac{\adC(1+z)^3+\abr(1+z)^{1/2}}{H(0)\Or^{1/2}}\right)^{1/2}\nonumber\\
&\approx \left(0.029\left(\frac{1+z}{2\times 10^6}\right)^3+4.18\times
  10^{-4}\left(\frac{1+z}{2\times 10^6}\right)^{1/2}\right)^{1/2}
,\label{xH}
\end{align}
$\xc$ and $\xH$ are plotted in Fig. \ref{xrates}. We note that $\xc<<1$ in
the redshift range of interest. This is consistent with assumptions made in
the derivation. Also,  since $\xc\sim 0.01$ in the redshift range of
interest and gaunt factors are slowly varying functions of frequency and
temperature, we can assume $\gdC\approx
\gdC(0.01)=1.005$ and $\gbr\approx 2.99$. 

{We should emphasize that
the sole purpose of finding an accurate solution for the spectrum at low frequencies
is to calculate precisely the photon emission/absorption due to
bremsstrahlung and double Compton. In particular, only the spectrum at
$\xe\gtrsim 0.1$ is frozen-in at $x\ll 10^5$. The low frequency spectrum
continues to be affected by bremsstrahlung  (at smaller redshifts, during
recombination and after), which tries to bring the spectrum in equilibrium
with the electrons, which are cooler than the radiation due to adiabatic
expansion \citep{cs2011,ksc2011}. Furthermore, most of the photons are created around
the critical frequency $\xc$. The double Compton gaunt factor at the critical
frequency $\gdC(\xc)$ deviates from the value $1.005$ by less than $0.5\%$
in the interesting redshift range of $10^5<z<10^7$. The bremsstrahlung
gaunt factor  has a maximum deviation of $\sim 7\%$ from $2.99$ in the same
redshift range, but since it is only a small correction to the dominant
double Compton process, the error in the final solution for $\mu$ evolution
is  small. Thus the assumption of constant gaunt factor is an
excellent one for the present problem and is further justified by a
comparison of the analytic and numerical solutions.}

We can now use the solution Eq. \eqref{xc} to evaluate the integral
Eq. \eqref{ndtint} (ignoring the $\xe$ dependence of gaunt factors),
\begin{align}
\frac{\id }{\id t}\ln\left(\frac{N}{N_{\gamma}}\right)&\approx\frac{\left(\KdC+\Kbr\right)}{I_2}\int\id
\xe\frac{\muc e^{-\xc/\xe}}{\xe^2}=\frac{\muc}{I_2}\frac{\KdC+\Kbr}{\xc}=\frac{\muc}{I_2}\left[\left(\KdC+\Kbr\right)\KC\right]^{1/2}.
\label{ndtintsol}
\end{align}
Equations \eqref{edt} and \eqref{ndt} along with the above solution give the following
equation for the evolution of chemical potential $\mu=\muc$ at $x>1$,
\begin{align}
\frac{\id \mu}{\id
  z}=\frac{C}{(1+z)H}\left[\mu\left[\left(\KdC+\Kbr\right)\KC\right]^{1/2}-B\frac{\SE}{\SQ}+\frac{4 B}{3}\left.\frac{\SN}{\Sn}\right|_{\rm Extra}\right],\label{mueq}
\end{align}
where $C=0.7768$, $B=1.803$. $\left.\frac{\SN}{\Sn}\right|_{\rm Extra}$ are
the extra photons injected from processes other than the low frequency bremsstrahlung and
double Compton photons calculated above, for example, from  the same source which
injects energy.  We  assume this extra term to be negligible in the rest
of the paper. The solution of Eq. \eqref{mueq} at $z=0$ is  given by
\begin{align}
\mu(0)=\mu(z_i)e^{-\mT(z_i)}+CB\int_{z_{\rm min}}^{z_i}\frac{\id
z}{(1+z)H}\left( \frac{\SE}{\SQ}-\frac{4 }{3}\left.\frac{\SN}{\Sn}\right|_{\rm Extra}\right)e^{-\mT(z)}\label{musol}
\end{align}
where $z_{\rm min}\approx 5\times 10^4$, $z_i$ is the initial/maximum energy
injection redshift and  we have defined the effective \emph{blackbody
  optical depth} (which is frequency independent at $x\gtrsim \xc$)

\begin{align}
\mT(z)&=\int_0^{z}\id z' \frac{C
    \left[\left(\KdC+\Kbr\right)\KC\right]^{1/2}}{(1+z')H}\nonumber\\
&\approx
\left[\left(\frac{1+z}{1+\zdC}\right)^5+\left(\frac{1+z}{1+\zbr}\right)^{5/2}\right]^{1/2}+\epsilon
\ln\left[\left(\frac{1+z}{1+z_{\epsilon}}\right)^{5/4}+\sqrt{1+\left(\frac{1+z}{1+z_{\epsilon}}\right)^{5/2}}\right],\label{optold}
\end{align}
where
\begin{align}
\zdC&=\left[\frac{25 \Or H(0)^2}{4 C^2\aC\adC}\right]^{1/5}=1.96\times
10^6\nonumber\\
\zbr&=\left[\frac{25 \Or H(0)^2}{4 C^2\aC\abr}\right]^{2/5}=1.05\times
10^7\nonumber\\
z_{\epsilon}&=\left[\frac{\abr}{\adC}\right]^{2/5}=3.67\times
10^5\nonumber\\
\epsilon&=\left[\frac{4 C^2\abr^2\aC}{25 \adC\Or H(0)^2}\right]^{1/2}=0.0151,
\end{align}
$\Or$ is the total radiation density parameter, and $H(0)$ is the Hubble
constant today. It is interesting to note that in the absence of double
Compton scattering we would have $\zbr\approx 6\times 10^6$. The
presence of double Compton increases the critical frequency $\xc$ from its
bremsstrahlung only value, and thus reducing the bremsstrahlung emission.
{It is also straightforward to calculate, if needed,  the chemical potential
  at any intermediate redshift $z'>z_{\rm min}$ using Eq. \eqref{musol} by replacing
  $\mathcal{T}(z)$ (and similarly for $\mathcal{T}(z_i)$) with
  $\mathcal{T}(z)-\mathcal{T}(z')$ and also replacing the 
  lower limit $z_{\rm min}$ in the integral with $z'$,
\begin{align}
\mu(z')=\mu(z_i)e^{-\left[\mT(z_i)-\mT(z')\right]}+CB\int_{z_{\rm '}}^{z_i}\frac{\id
z}{(1+z)H}\left( \frac{\SE}{\SQ}-\frac{4 }{3}\left.\frac{\SN}{\Sn}\right|_{\rm Extra}\right)e^{-\left[\mT(z)-\mT(z')\right]}\label{musolz}
\end{align}

The solution given in
  Eq. \eqref{xc} corresponds to Eq. (15) in \citep{sz1970} but including the
  double Compton process. Similarly, the solution in Eq. \eqref{optold} generalizes 
Eq. (20) of \citep{sz1970}. The dominant term ($\zdC$)in Eq. \eqref{optold} is due
to the double Compton process with the bremsstrahlung term ($\zbr$) providing a
small but important correction.}
\section{Improved solution by approximating non-stationarity using previous
  solution}
We will see below that the solution arrived at in the previous section underestimates the photon
production. It turns out that the stationary solution, which is normalized
at high frequencies, underestimates the chemical potential at small
frequencies where most of the photons  are being produced/absorbed. 
{We  find below the correction for the normalization of chemical potential,
which enables  us to improve the formula for blackbody optical depth
Eq. \eqref{optold} in the solution Eq. \eqref{musol}.
The result of the  computations,  using the new analytic formula
Eq. \eqref{optnew} for the
blackbody optical depth, deviates from numerical solution by less than
$1\%$.}

A
very simple correction to the normalization can be arrived at as follows.
An immediate improvement over the  solution of \citep{sz1970} is possible by
approximating the non-stationarity in Eq. \eqref{stateq} using the solution
Eq. \eqref{mueq} (ignoring the energy injection term)\footnote{The energy
  injection term will add a inhomogeneous term to the homogeneous equation for
  $\mu(x)$. The effect of this term is to change the overall normalization
  $\muc$ without significantly affecting the shape of the spectrum. This term
  can therefore be neglected for the purpose of calculating the photon
  creation/absorption.}
\begin{align}
\frac{\partial n}{\partial t}\approx \frac{-1}{\xe^2}\frac{\partial
  \mu}{\partial t}\approx \frac{C\mu\left[\left(\KdC+\Kbr\right)\KC\right]^{1/2}}{\xe^2}
\end{align}
Note that the time derivative of temperature in Eq. \eqref{kineticeq} can be neglected at small
frequencies as its effect
is suppressed by a factor of $\xe$ with respect  to the term with the time derivative of $\mu$.
Equation \eqref{stateq} with the above approximation for the non-stationary term gives
Bessel's equation
\begin{align}
\frac{\id
  }{\id \xe}\xe^2\frac{\id \mu}{\id
      \xe}-\left(\frac{\xc^2}{\xe^2}-C\xc \right)\mu&=0\label{stateq2}
\end{align}
The solution is given in terms of modified Bessel function of second kind $K_{\nu}(x)$,
\begin{align}
\mu(\xe)&=A\mu_c\sqrt{\frac{2}{\pi}}\sqrt{\frac{\xc}{\xe}}K_{0.5\sqrt{1-4C\xc}}(\xc/\xe).\label{mux2}
\end{align}
{This result provides a more precise dependence of $\mu$ on frequency
compared to Eq. \eqref{xc}.}
Choosing normalization to give $\mu(0.5)\approx \muc$ at $\xe=0.5$ gives
$A=1.007 + 3.5\xc$, where this fit is accurate for $5\times
10^{-3}<\xc<2\times 10^{-2}$. This fit thus covers all the interesting
range for critical frequency $\xc$ (see Fig. \ref{xrates}). This choice of
normalization frequency ($\xe=0.5$) provides a good fit to the numerical
solution.

\begin{figure}
\includegraphics[width=14cm]{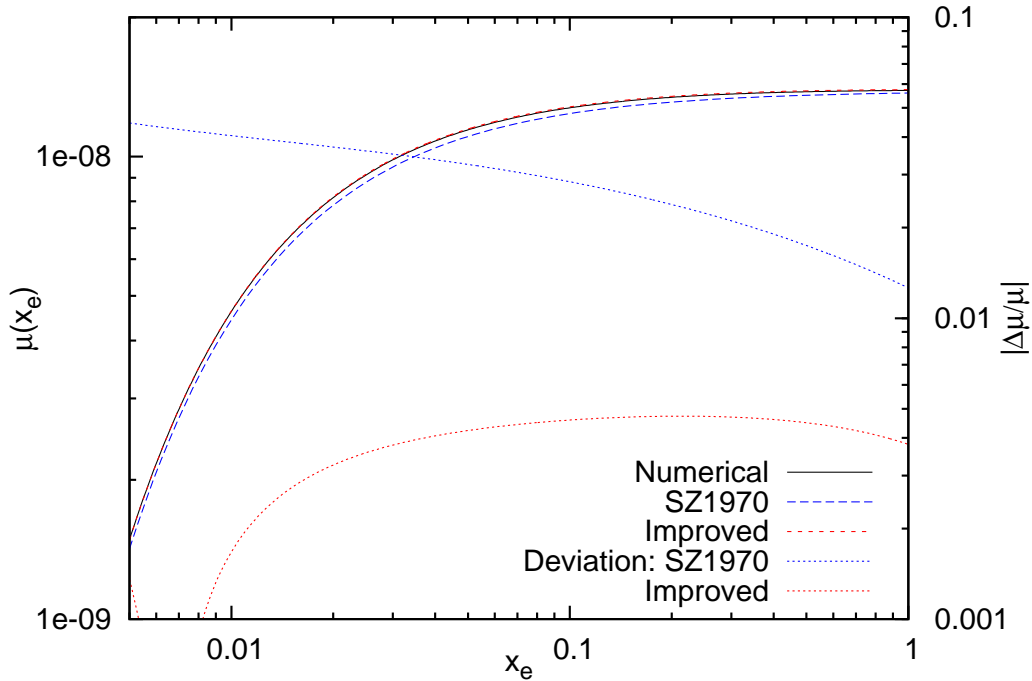}
\caption{\label{muxfig}Snapshot of the numerical solution, original solution Eq. \eqref{xc}
  marked SZ1970  and improved
solution Eq. \eqref{mux2} at $z =3.48\times 10^6$ (chosen randomly) with the initial
$\mu=10^{-5}$ at $z=5\times 10^6$ including both double Compton and bremsstrahlung photon
production and no additional energy injection. The critical frequency at
$z=3.48\times 10^6$ is $\xc=0.0114$. Both the analytical solutions are plotted
with the 
high frequency distortion $\muc=1.39\times 10^{-9}$. The numerical and
improved solutions are indistinguishable in the figure while the original
solution underestimates  $\mu(\xe)$ at low frequencies. 
}
\end{figure}

{  The  
normalization frequency is chosen so that (i) it is $<1$, since this is the
assumption made in deriving the analytic solution and (ii) it is also large enough
so that $\mu\approx {\rm constant}$. Since  we only want to
use this solution to calculate the total photon emission, it need only be
accurate at $\xe\lesssim 1$. The only requirement at $\xe>1$ is that its
contribution to the photon emission/absorption should be negligible at
$\xe\gg 1$ and that it should be approximately constant around $\xe=1$, as
expected from a correct solution. We should also point out that $\mu(\xe)$
decreases with increasing $\xe$ at $\xe\gg 1$ for the solution in
Eq. \eqref{mux2} and 
$\mu(\xe\rightarrow\infty)=0$. Thus our solution satisfies the requirements outlined
above.  Obviously it cannot be normalized at $\xe=\infty$,  as was done
with the original solution Eq. \eqref{xc}. The normalization must be done by comparison
with the numerical solution, taking into account the assumptions made in
arriving at this solution, resulting in our choice of
$\xe=0.5$.\footnote{{Since $\mu(\xe)$ is approximately constant around
  $\xe=0.5$ (variation in the analytic solution is less than $1\%$  for $0.4<\xe<1$), the exact value of normalization frequency is not important,
  and we get similar precision if we choose to normalize at a slightly
  different frequency around $\xe=0.5$, for example at $\xe=0.6$. We should
also mention that the numerical solution also shows a tiny decrease ($\sim 1\%$
from $\xe=1$ to $\xe=100$) in the
chemical potential at $\xe>1$ because of the increasing efficiency of
the recoil effect at high frequencies.}} We show a snapshot of the  numerical solution (chosen at random)
  at $z=3.48\times 10^6$, original solution Eq. \eqref{xc} and improved
solution Eq. \eqref{mux2} in Fig \ref{muxfig}. The critical frequency at
this redshift is $\xc=0.0114$. Needless
to say that the shape of the spectrum is well described by
our solution at all redshifts and we have chosen a random snapshot in
Fig. \ref{muxfig} for illustration. The final justification for all our
assumptions and approximations is of course given by a comparison of the
final 
numerical and analytic solutions 
 for the evolution of the chemical potential with redshift as described below.}

We can now use our improved solution to calculate the photon production
rate 
\begin{align}
\frac{\id }{\id t}\ln\left(\frac{N}{N_{\gamma}}\right)&\approx\frac{\left(\KdC+\Kbr\right)}{I_2}\int_0^{\infty}\id
\xe\frac{\mu(\xe) }{\xe^2}\nonumber\\
&\approx\frac{\muc}{I_2}(\KdC+\Kbr)(1.007+3.5
\xc)\left(\frac{1}{\xc}-C\ln(2)\right)\nonumber\\
&\approx \frac{\muc}{I_2}\left[1.007\left[\left(\KdC+\Kbr\right)\KC\right]^{1/2}+2.958\left(\KdC+\Kbr\right)\right].
\label{ndtintsolnew}
\end{align}
Proceeding as before we get improved formula for blackbody optical depth,
\begin{align}
\mT(z)&=\int_0^{z}\id z' \frac{1.007C
    \left[\left(\KdC+\Kbr\right)\KC\right]^{1/2}+2.958C\left(\KdC+\Kbr\right)}{(1+z')H}\nonumber\\
&\approx
1.007\left[\left(\frac{1+z}{1+\zdC}\right)^5+\left(\frac{1+z}{1+\zbr}\right)^{5/2}\right]^{1/2}+1.007
\epsilon
\ln\left[\left(\frac{1+z}{1+z_{\epsilon}}\right)^{5/4}+\sqrt{1+\left(\frac{1+z}{1+z_{\epsilon}}\right)^{5/2}}\right]\nonumber\\
&+\left[\left(\frac{1+z}{1+\zdC'}\right)^3+\left(\frac{1+z}{1+\zbr'}\right)^{1/2}\right],\label{optnew}
\end{align}
where we have defined 
\begin{align}
\zdC'&=\left[\frac{3 \Or^{1/2} H(0)}{2.958 C\adC}\right]^{1/3}=7.11\times
10^6\nonumber\\
\zbr'&=\left[\frac{ \Or^{1/2} H(0)}{5.916 C\abr}\right]^{2}=5.41\times
10^{11}\nonumber\\
\end{align}
The improved solution for evolution of $\mu$  is still given by the original
equation \eqref{musol} on substituting   the improved  optical depth given by
Eq. \eqref{optnew}. 

{The improved solution has a broad  region of validity and covers
  the entire redshift range of interest.
The  physics  used to derive Eq.  \eqref{optnew} is applicable for  
redshifts  $z\lesssim 8 \times 10^7$. Thus we can use the blackbody
optical depth, Eq.  \eqref{optnew},  for redshift interval $10^5 < z <
8\times  10^7$, the upper limit is well
behind blackbody surface at $z\sim   2 \times 10^6$. At $z\approx 8\times
10^7$  the number density of positrons becomes comparable to the number
density of electrons/baryons due to
  pair production. At higher redshifts, the number density of electrons and
  positrons, and thus the rates of Compton scattering, double Compton and 
  electron-electron and electron-positron bremsstrahlung, increase
  exponentially with increasing redshift. Thus the blackbody optical depth, $\mathcal{T}$ also
  starts increasing exponentially instead of a  power law as in our
  solution and  Eq.  \eqref{optnew} is no longer applicable. However, we already have $\mathcal{T}\sim 10^4$ at $z\sim 8\times 10^7$,
  and   creation of a distortion in photon spectrum is thus impossible at higher redshifts.}

\begin{figure}
\includegraphics[width=14cm]{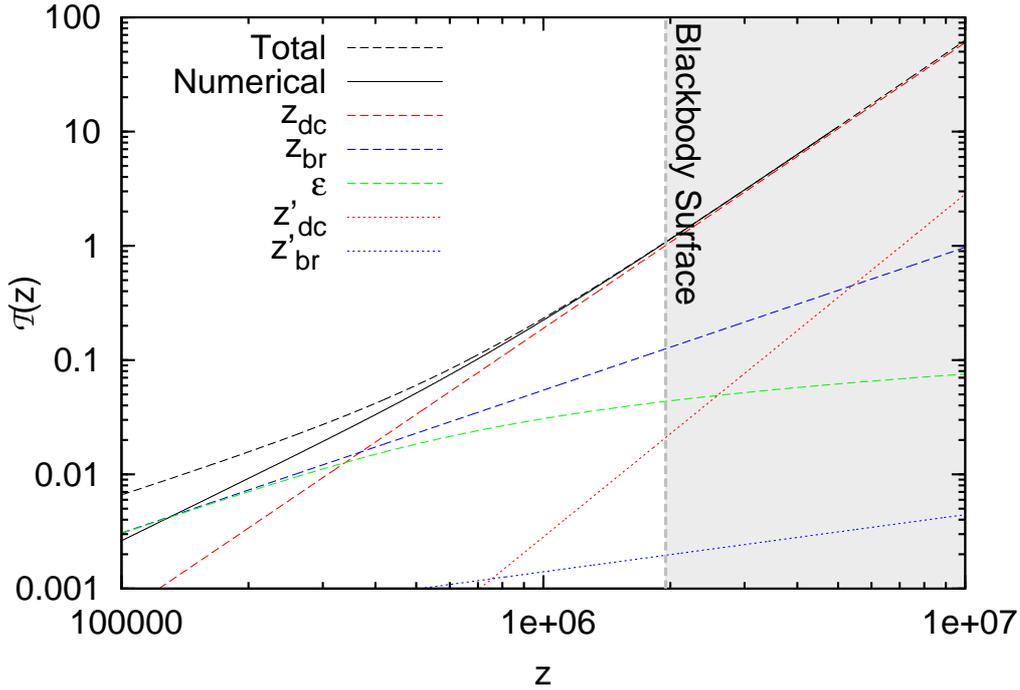}
\caption{\label{optfig}Blackbody optical depth as a function of redshift (independent of frequency)
  calculated using  numerical solution,   and improved
solution Eq. \eqref{optnew}. Individual terms in Eq. \eqref{optnew} are also
shown. At high redshifts double Compton terms dominate with the new double
Compton term $\zdC'$ also contributing. Bremsstrahlung term becomes
important at low redshifts. At high redshifts  our improved solution is
indistinguishable from the numerical solution. The new solution is an
excellent 
approximation to the numerical result  over the entire redshift of interest
 where the optical depth is greater than a few $\%$. 
}
\end{figure}

\begin{figure}
\includegraphics[width=14cm]{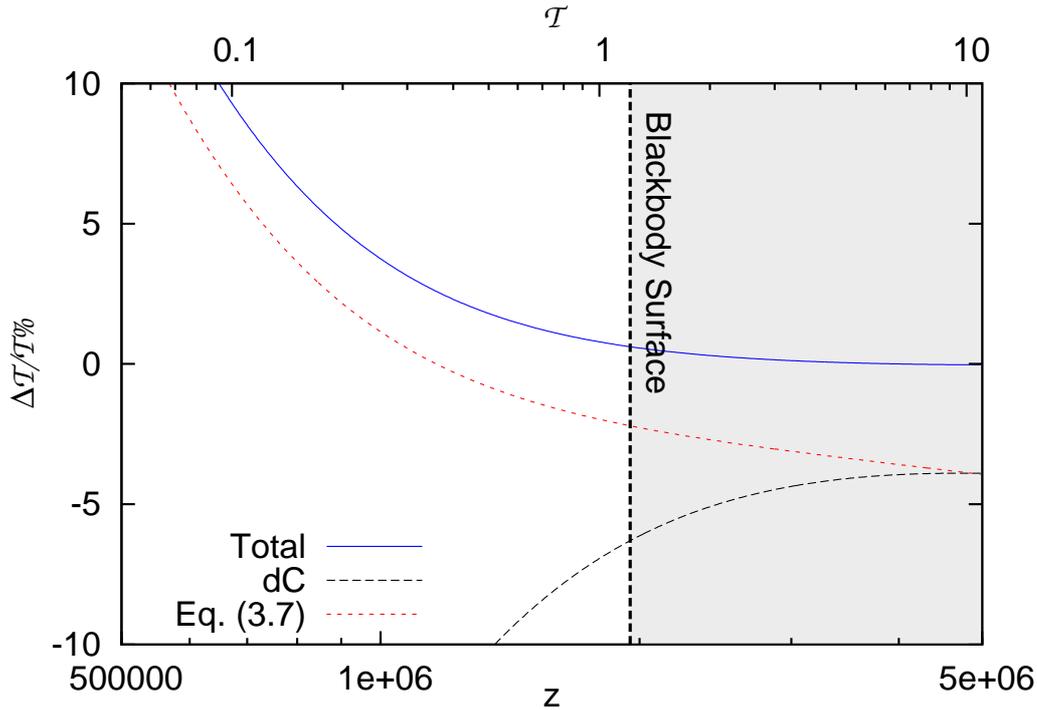}
\caption{\label{opterror} Deviation ($\%$) from the numerical solution in blackbody optical depth with respect to the
  numerical solution for the standard double Compton only analytic
  solution, Eq. \eqref{optold}, which includes both bremsstrahlung and
  double Compton using the method of \citep{sz1970}
and our new solution. The error in $\mathcal{T}$ at low redshifts does not
have a significant 
effect on the final spectrum. The  error in visibility for our improved
solution is better than
$1\%$ at all redshifts and shown in Fig. \ref{visfig}.}
\end{figure}

\begin{figure}
\includegraphics[width=14cm]{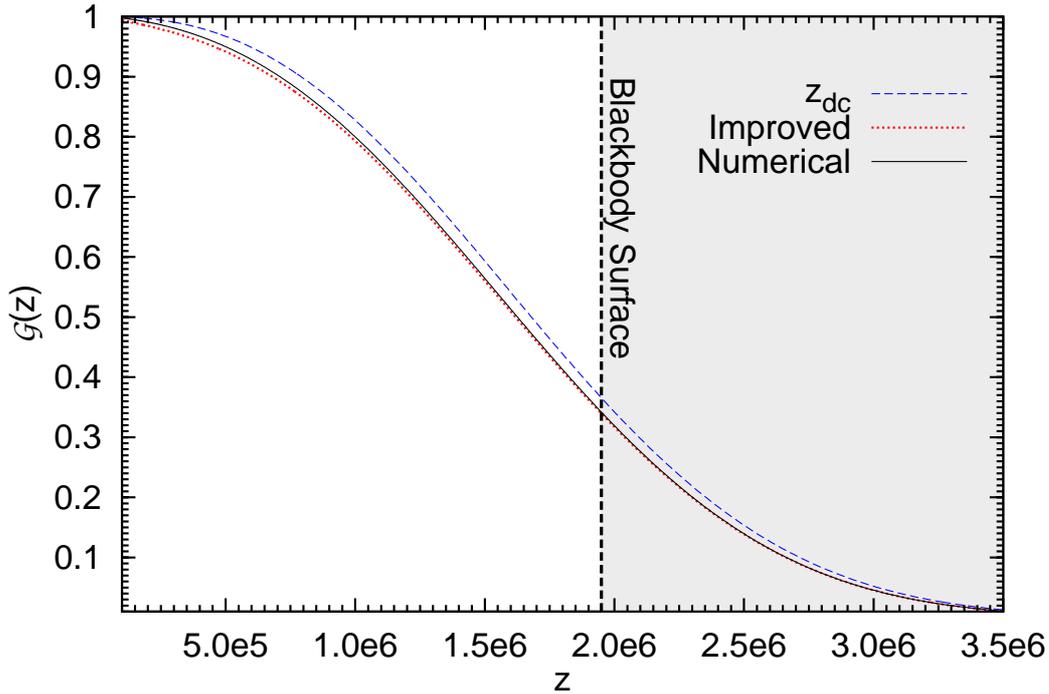}
\caption{\label{visfigzoom}{The characteristics of the blackbody photosphere as
  given by the blackbody visibility} $\mathcal{G}\equiv e^{-\mathcal
 {T}}$ for analytic
solution Eq. \eqref{optnew},  for the double Compton only solution $\zdC$ and the
numerical result. We have introduced  \emph{blackbody surface} as the boundary
where the blackbody optical depth $\mathcal{T}$=1.}
\end{figure}

\begin{figure}
\includegraphics[width=14cm]{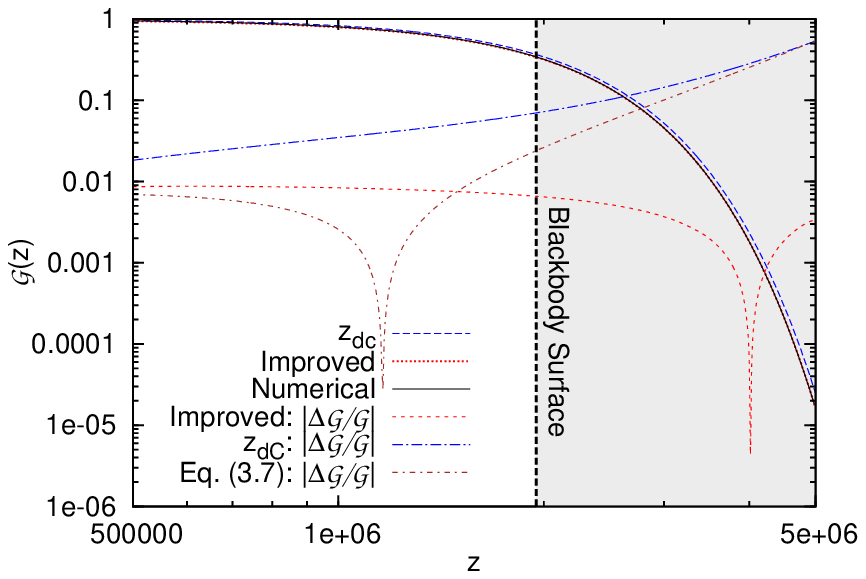}
\caption{\label{visfig} Same as Fig. \ref{visfigzoom} but going to higher
  redshifts and also showing  
the  errors, relative to the numerical solution, for the different analytic
solutions. Deviations from the numerical solution is also shown for Eq. \eqref{optold} based on the method of
\citep{sz1970} but including both the double Compton and
bremsstrahlung. The errors are negative at high redshifts and positive at
low redshifts with a spike where they change sign.}
\end{figure}

We plot the optical depth for the numerical solution,  the total improved solution
 as well as the individual terms in Eq. \eqref{optnew} in Fig. \ref{optfig}.
At high redshifts, double Compton terms dominate with the new double
Compton term $\zdC'$ also contributing. Bremsstrahlung term becomes
important at low redshifts. At high redshifts,  our improved solution is
indistinguishable from the numerical solution.  As blackbody optical
 depth becomes small the quasi-static assumptions made in arriving at the
 analytic solution also breakdown and the error grows. The new solution is
 however an
excellent 
approximation to the numerical result, and definitely better than the
double Compton only result, over the entire redshift range  of interest,
 where the optical depth is greater than a few $\%$.
The deviations from the numerical result for the double Compton only formula and our new result are
plotted in Fig. \ref{opterror}. The analytic solution overestimates the
photon production at low redshifts. The reason becomes clear by looking at
$\xH$ in Fig. \ref{xrates}. Photon production at $\xe>\xH$ would be
suppressed since the photon production rate is smaller than the expansion
rate. At high redshifts, $\xH\gtrsim 1$ and the error introduced by
including $\xe>\xH$ is negligible since photon production is negligible at
these frequencies anyway. At low redshifts, $\xH$ becomes less than
unity and starts approaching $\xc$ and Eqs. \eqref{ndtintsol} and \eqref{ndtintsolnew} overestimate photon
production. We show the \emph{blackbody visibility} factors\footnote{{This
  is really the visibility of distortions. When the visibility is small,
  the distortions are not visible, and when the visibility is unity,
  distortions survive and are visible today.}} 
$\mathcal{G}\equiv e^{-\mathcal{T}}$ in Figs. \ref{visfigzoom} and \ref{visfig} for analytic
solution Eq. \eqref{optnew}, for the double Compton only term $\zdC$ and the
numerical result. The accuracy of blackbody visibility is better than $1\%$
with the new solution Eq. \eqref{optnew}.


\section{Examples from standard cosmology}
\subsection{Upper limit to energy release after BBN  and before recombination}
{In standard model of cosmology, we can get constraints on energy density in
radiation from two distinct and very precise observables. The first is the
deuterium abundance, which gives the baryon number  to photon number
ratio $\eta=(5.7\pm 0.3)\times 10^{-10}$ during
primordial nucleosynthesis at $4\times 10^8 \gtrsim z\gtrsim 4\times 10^7$ \citep{fields2001,iocco2009}.  The second is the
measurement of CMB anisotropies, which constrains the baryon to photon
ratio $\eta=(6.18\pm 0.15)\times  10^{-10}$ \citep{wmap7} during recombination at $z\approx 1100$. The fact that these
two values of baryon to photon ratio are almost identical, with  small
error bars means that we do not have arbitrary freedom in adding energy to
CMB, for example, with the
introduction of new physics. In fact any addition of energy/entropy to
CMB between primordial nucleosynthesis and recombination cannot be more
than a small percentage ($\sim  7\%$ for CMB and BBN to be consistent within $2-\sigma$) of the already
existing  radiation energy density. COBE limit \citep{cobe} of $\mu\lesssim
9\times 10^{-5}$ implies that a $\sim 7\%$ energy can be added only at
redshifts $z\gtrsim 4.1\times 10^6$.  These limits also
justify our assumption of small distortions in the analytic
calculations. Any energy injection into photons from non-standard processes before
electron-positron annihilation, however, is unconstrained.}

\subsection{Electron-positron annihilation}
It is, of course, possible to add of order unity energy to radiation
before primordial nucleosynthesis.  {This happens in standard
cosmology during electron-positron annihilation
\citep{alpher53,peebles66,zeld66}, which more than doubles the energy
density of photons and increases their temperature by $\sim 40\%$.}
In the early stages of electron-positron annihilation, the
electrons/positrons far outnumber the photons. In this era, therefore,
electron-positron annihilation and electron-electron/electron-positron 
bremsstrahlung dominate the thermalization process. {In the very late
stages, when most of the electron-positrons have annihilated, Compton and double
Compton scattering 
are dominant, electron number is conserved and their density evolves
according to the
non-relativistic adiabatic law, and our analytic formulae become applicable.} {We, of course,
do not expect any observable $\mu$ distortion from electron positron
annihilation \citep{sz1970}.} It is still interesting to calculate the magnitude of the
distortion to demonstrate the effectiveness of double Compton scattering and
comptonization in restoring the equilibrium between matter and radiation.

At redshifts $z\lesssim 10^8$, most of the positrons have annihilated and
their number density falls below that of electrons. The number density of electrons is high
enough  (as a result of $\sim 10^{-9}$
asymmetry in matter anti-matter) to maintain the annihilation  rate much faster than the
expansion rate. We can, thus, use Saha equation to follow the positron number
density during the last stages of positron annihilation. Using the fact that the positron number
density is much smaller than the electron number density and that the
electron number density is unaffected by annihilation, we get for the positron
number density $n_{+}$,
\begin{align}
n_{+}&\approx\frac{\neql^2}{\Ne},
\end{align}
where the equilibrium (zero chemical potential) number density of electrons/positrons is 
\begin{align}
\neql&=\frac{2}{h^3}\left(2\pi \me \kB T\right)^{3/2}e^{-\frac{\me
      c^2}{\kB T}}
\end{align}
The rate of energy injection is  given by,
\begin{align}
\SE=H(1+z)\frac{2\me c^2}{\aR T^4}\frac{\id  n_{+}}{\id z}
\end{align}

\begin{figure}
\includegraphics[width=14cm]{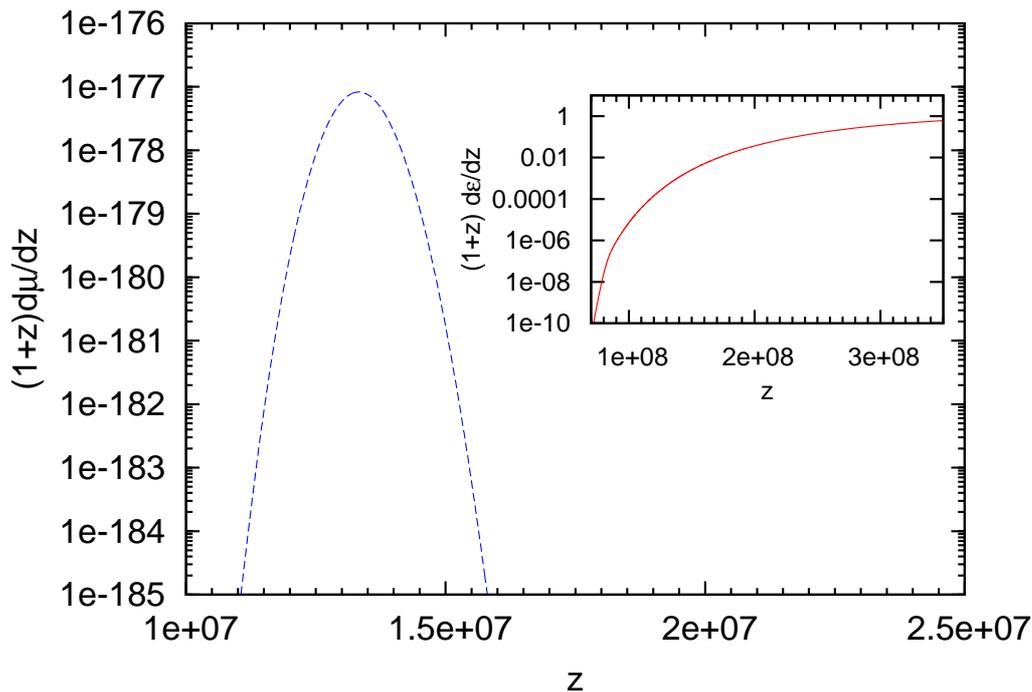}
\caption{\label{muex} Chemical potential $\mu$ from electron positron
  annihilation. The CMB blackbody spectrum is maintained at extraordinary
  precision of $10^{-178}$! Also shown, in the inset, is the actual rate of energy
  injection multiplied by $(1+z)$. At high redshifts ($z\gtrsim 10^8$) we have used entropy
  conservation to calculate the rate of heating.
}
\end{figure}

We have plotted the resulting $\mu$ injection rate multiplied by redshift,
$BC (1+z)\frac{2\me c^2}{\aR T^4} \frac{\id  n_{+}}{\id z}e^{-\mathcal{T}}$
in Fig. \ref{muex}. Visibility function suppresses the high redshift
contribution, while the exponentially decreasing positron number density
suppresses the low redshift contribution, giving the peak at $z\sim
1.3\times 10^7$. The
chemical potential from electron-positron annihilation is suppressed by an
astronomical factor of $10^{178}$! Thus it is impossible to create a
deviation from blackbody spectrum at high redshifts.

\subsection{Primordial nucleosynthesis}
Big bang  nucleosynthesis (BBN) at $z\sim 3\times 10^8$ results in binding of almost all neutrons into
helium ($\nHe/\nB\approx 6\times 10^{-2}$) along with the production of small amount of deuterium
($\nHtw/\nB\approx 2\times 10^{-5}$), helium-3
$\nHet/\nB\approx 8\times 10^{-6}$,  tritium $\nHt/\nB \approx 7\times
10^{-8}$, beryllium-7/lithium-7
($\nBes/\nB\approx 3\times 10^{-10}$),  lithium-6 ($\nLisi/\nB\approx
10^{-14}$) and trace amounts of heavier elements \citep[see ][ for a recent calculation]{serpico}.
We can get a rough estimate of the energy released during this main part of
nucleosynthesis by calculating the total binding energy of helium-4 produced.\footnote{We
  ignore the fact that some of the energy will be lost to neutrinos.} Thus
we have an energy release of $\Delta E/E\sim E_{bind}\nHe/E_{\gamma} \sim
6\times 10^{-9}$. The blackbody optical
depth at $z\sim 10^8$ is $\sim 10^5$ and we have the final  $\mu\sim 0$. However, tritium and
beryllium-7 survive for a long time before decaying into helium-3 and
lithium-7 respectively. {Although, the energy released in the decay
  of beryllium-7 and lithium-7 is much smaller than that released during
  helium production in BBN, the distortions are much larger because these
  decays happen in front of the blackbody surface, when the blackbody visibility is almost
  unity. Also, the energy density released is proportional to the number
  density of  beryllium-7 and lithium-7, which  has
  decreased as $(1+z)^3$ compared to the $(1+z)^4$ decrease of the radiation energy
  density, thus giving a larger  $\Delta E/E$ than if the decay had
  happened at the same time as the main BBN.}  Tritium has a half life of $12.32$ years. It, therefore, decays at $z\sim
2.5\times 10^5$   to helium-3 releasing an electron with average energy
$5.7$KeV. Most of this energy release happens at $z\gtrsim 10^5$ and causes a $\mu$-distortion with
$\mu=2\times 10^{-15}$. 

Neutral beryllium atom decays by electron capture with a  half life of 53.2
days. Fully ionized beryllium in the low density plasma in the early
Universe is however stable. It has  to wait for $\approx 800$ years
until $z\approx 3\times 10^4$ when it recombines to hydrogen like beryllium. The
recombined beryllium can now capture the orbital electron and decay to
lithium-7 with a half-life of 106.4 days, which is twice the half-life of a
fully recombined beryllium \citep{ks2011}. $89.6\%$ of the decays go to
the ground state of lithium-7 and most of the energy released is carried away
by neutrinos, which would  appear today as a narrow line in the cosmic neutrino
spectrum. $10.4\%$ of beryllium  decays into an excited state of lithium. About half of the total decay energy in this case also is lost to neutrinos
forming a second lower energy line in the cosmic neutrino spectrum.
The excited lithium nucleus then de-excites, almost immediately, to the ground state, emitting a $Q=477.6$ KeV photon, which
delivers most of its energy to plasma by Compton scattering on electrons
(recoil effect). The Compton $y$ parameter at $z=30000$ is $0.04$, which
lies intermediate between  pure $y$ and $\mu$ type eras. The
heating, therefore, results in a distortion intermediate between the $y$
and the $\mu$ type
distortions of magnitude (using formula for $y$-type distortion) $\approx
(1/4)\Delta E/E \approx (1/4)0.104 Q \nBes
/E_{\gamma}\sim 10^{-16}$.

\subsection{Dark matter annihilation}
A natural and favored candidate for dark matter is a weakly interacting massive
particle (WIMP) with several candidates in high energy theories beyond the
standard model \citep{dm}. A very attractive feature of WIMP 
is that  if they have  weak scale interactions then the correct amount
of dark matter (which is close to the critical density) observed today can be thermally produced in the early Universe.
This is remarkable since a priori there is no reason to suspect any
relation between the weak scale
interactions and the present critical density of the Universe and this
coincidence is sometimes
referred to as the \emph{WIMP miracle}. For the thermally produced WIMPs, the
dark matter density is related to the velocity averaged cross section as
\begin{align}
\left<\sigma v\right>&\approx \frac{3\times 10^{-27}}{\Odm h_0^2}{\rm cm^3s^{-1}},
\end{align}
where $h_0\approx 0.702$ is the Hubble parameter and $\Odm$ is the dark
matter density as a fraction of critical density today. The above values for
annihilation cross sections are of similar order of magnitude as the
current upper
limits from  Fermi-LAT experiment probing dark matter annihilation in the local
Universe \citep{fermi2,fermi1,hutsi}. {The actual annihilation cross section
can, of course, be much smaller than the Fermi upper limits.} {Energy
released from dark matter
annihilation  also changes recombination history \citep{ss1985} and it can
thus be constrained through its effect on the CMB power spectrum \citep{ck2004,pf2005}
and recombination spectrum \citep{c2010}. Effect of dark matter
annihilation also changes the abundance of elements produced during BBN
\citep{bbn1,bbn2}; these constraints are complementary but less stringent compared to
the
current 
CMB constraints.}

Initially, dark matter is in thermal equilibrium with other constituents of
the Universe, and is being continuously created and annihilated. As the Universe cools and expands, these interactions freeze out
and thereafter the dark matter number is  conserved. However, a small number of  residual
annihilations keep happening throughout the history of the Universe. 
The rate of energy released into CMB from  these residual
annihilations\footnote{We assume self annihilating Majorana particles. For
  Dirac particles the energy release would be smaller by a factor of 2.} is given
by,
\begin{align}
 \SE&=f_{\gamma}\frac{\mdm c^2n_{dm}^2\left<\sigma
    v\right>}{a T^4}\nonumber\\
&\approx 1.4\times 10^{-29}(1+z)^2f_{\gamma}\left(\frac{10{\rm
      GeV}}{\mdm}\right)\left(\frac{\Odm h_0^2}{0.105}\right),
\end{align}
where $\mdm$ is the mass of the dark matter particle and $f_{\gamma}$ is the
fraction of energy that goes into particles with electromagnetic
interactions, and is deposited in the plasma.

\begin{figure}
\includegraphics[width=14cm]{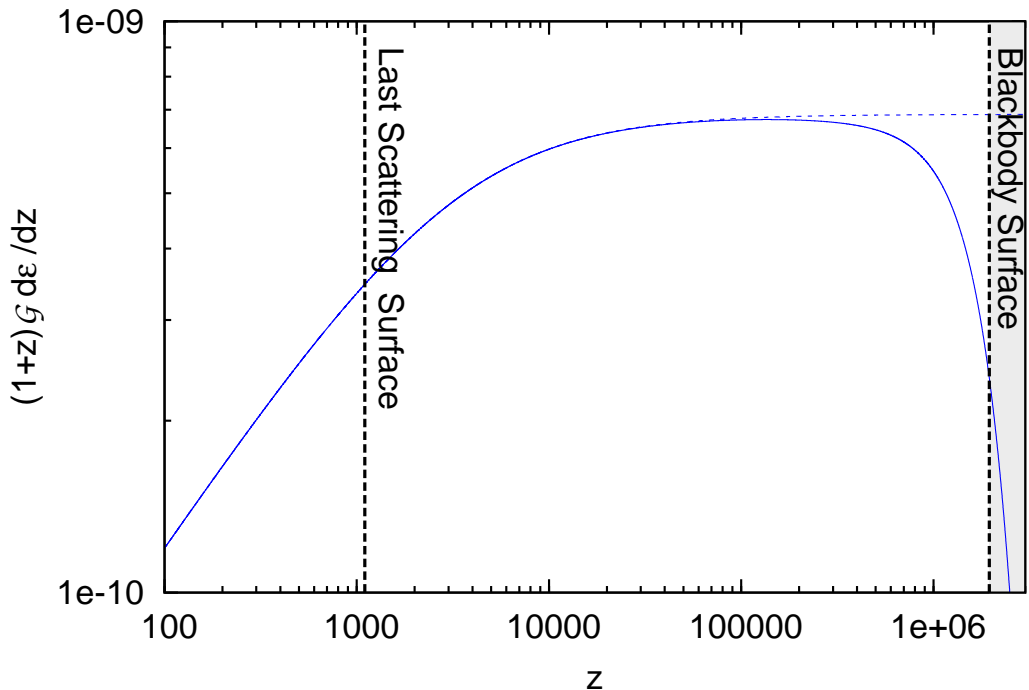}
\caption{\label{dmfig} Energy injection from dark matter annihilation for a
  $10~{\rm GeV}$ WIMP with $f_{\gamma}=1$ (solid line). The
  exponential suppression at high redshifts is because of the decrease in
  visibility, $e^{-\mathcal{T}}$. Dashed line shows the energy injection 
  $(1+z)\frac{\id \SQ}{\id z}$ without the visibility factor.
}
\end{figure}

We plot the rate of energy release into the CMB   
\begin{align}
(1+z)\mathcal{G}\frac{\id \SQ}{\id z}&= \SE\frac{e^{-\mathcal{T}}}{H}
\end{align}
in Fig. \ref{dmfig} for the fraction of energy going into the plasma\footnote{In
  general we expect some energy to be lost to neutrinos and other dark
  particles and therefore $f_{\gamma}$ would be less than unity.}
$f_{\gamma}=1$. Hubble rate is  proportional to $(1+z)^2$ during
radiation domination, which gives us the flat portion of the curve. Energy
release rate decreases faster than the expansion rate $\propto (1+z)^{3/2}$
during matter domination, giving the low redshift declining tail in the
plot. The total $\mu$ distortion (using energy release from
$z> 5\times 10^4$) is $\mu\approx 3\times 10^{-9}$. The $y$- type
distortion from energy release $z\lesssim 5\times 10^4$ is $y\approx 5\times
10^{-10}$, these distortions were also calculated by  \citep{cs2011}. These numbers are of similar order of magnitude as the
distortions from Silk damping and Bose-Einstein condensation of CMB
discussed in the next two sections. COBE constraint of $\mu<9\times 10^{-5}$
\citep{cobe} constrains WIMP mass to be $\mdm >0.3f_{\gamma}{\rm MeV}$ while
PIXIE \citep{pixie} would be able to constrain up to $\mdm \sim 3f_{\gamma}{\rm GeV}$.

\subsection{Silk damping}
Sound waves are excited in the primordial baryon-electron-photon plasma  by
primordial perturbations.  They  decay on small scales  because
of shear
viscosity, with thermal
conduction also becoming important  near the time of recombination.  This damping of primordial perturbations was first calculated by
Silk \citep{silk} including only thermal conduction. The full
calculation, including both shear viscosity and thermal conduction and also
including
the effects of photon polarization was done by Kaiser \citep{kaiser}. Silk damping transfers energy from
sound waves to the average CMB spectrum, resulting in effective energy
injection into CMB \citep{sz1970b,daly1991,hss94,cs2011,cks2012}. Microscopically, shear viscosity and thermal conduction arise
due to the diffusion of photons which are repeatedly scattered by
the electrons. This diffusion of photon results in mixing of blackbodies from
different phases of the sound waves on diffusion scales. $2/3$ of the
dissipated 
energy in sound waves just increases the average temperature of CMB while
$1/3$ results in the spectral distortions of $\mu$ and $y$ type. Depending on
the primordial perturbation power spectrum at these very small scales of comoving wavenumbers
$50\lesssim k\lesssim 10^4$, the $\mu$ distortion can be in the range
$10^{-8}-10^{-10}$ \citep{cks2012} for the parameter space allowed in the standard cosmological
model \citep{wmap7}. We refer to  \citep{cks2012} for a detailed discussion,
including  fitting formulae for spectral distortions from  adiabatic
initial conditions, and constraints from the future experiments on initial power
spectrum spectral
index and its running.

\subsection{Bose-Einstein condensation of CMB}
After the epoch of electron-positron annihilation, electrons and baryons
are non-relativistic and cool adiabatically (with adiabatic index $5/3$) as
a result of the expansion of the
Universe, $\Te\propto (1+z)^2$. Radiation
(photons) has adiabatic index $4/3$ and cools slower than baryons,
$T_{\gamma}\propto (1+z)$ \citep{zks68}. Comptonization however is very efficient before
recombination and efficiently transfer energy from photons to
electrons/baryons, keeping them at same temperature as photons. This cooling
of CMB \citep{cs2011}, along with thermalization from comptonization, results in
Bose-Einstein condensation of CMB \citep{ksc2011}. The photons thus move from high to low
frequencies where they are efficiently destroyed by bremsstrahlung (and at
high redshifts also by double Compton scattering). Since the amount of cooling
is small, linear theory for small distortions applies. The resulting
distortions have  the same shape as that caused by heating of CMB in
previous examples, but with opposite sign. Thus we have negative $\mu$ and
negative $y$ distortions which partially cancel the distortions  due to
dark matter annihilation and Silk damping. Surprisingly, the $\mu$ (and $y$)
distortions have a magnitude which is similar to those from dark matter
annihilation and Silk damping. A comparison of $\mu$ distortions from
Bose-Einstein condensation as well as all previous examples is presented in
Table \ref{tbl}. {We also show comparison of $y$-type distortions in Table
\ref{tbl2}. $y$-type distortions are dominated by the low redshift
contributions, during and after reionization, from the intergalactic medium and
clusters. Early universe physics is therefore difficult to constrain using
the $y$-type distortions.}

\begin{table}
\begin{tabular}{|c|c|}
\hline
 Process  & $ \mu$ \\
\hline
\color{blue} electron-positron annihilation & \color{blue}$  10^{-178}$\color{black}\\ 
\color{blue} BBN tritium decay & \color{blue}$ 2\times  10^{-15}$\color{black}\\ 
\color{blue} BBN $^7{\rm Be}$ decay & \color{blue}$ 10^{-16}$\color{black}\\ 
\color{blue} WIMP dark matter annihilation & \color{blue}$ 3\times
10^{-9}f_{\gamma}\frac{10{\rm GeV}}{\mdm}$\color{black}\\ 
\color{blue} Silk damping & \color{blue}$ 10^{-8} - 10^{-9}$\color{black}\\ 
\color{red}Adiabatic cooling of matter and & \color{black}\\ 
\color{red}Bose-Einstein condensation & \color{red}$ -2.7\times 10^{-9}$\color{black}\\ 
\hline
\end{tabular}
\caption{\label{tbl}Census of energy release and $\mu$ distortions  in
  standard cosmological model.  The negative distortion from adiabatic cooling of
matter is shown in red.} 
\end{table}

\begin{table}

\begin{tabular}{|c|c|}
\hline
 Process  & $y$ \\
\hline
\color{blue} WIMP dark matter annihilation & \color{blue}$ 6\times
10^{-10}f_{\gamma}\frac{10{\rm GeV}}{\mdm}$\color{black}\\ 
\color{blue} Silk damping & \color{blue}$ 10^{-8} - 10^{-9}$\color{black}\\ 
\color{red}Adiabatic cooling of matter and & \color{black}\\ 
\color{red}Bose-Einstein condensation & \color{red}$ -6\times 10^{-10}$\color{black}\\ 
\color{blue}Reionization & \color{blue}$  10^{-7}$\color{black}\\ 
\color{blue}Mixing of blackbodies: CMB $\ell\ge 2$ multipoles & \color{blue}$  8\times 10^{-10}$\color{black}\\ 
\hline
\end{tabular}
\caption{\label{tbl2}Census of energy release and $y$ distortions  in
  standard cosmological model. We also give the value of $y$-type distortion
  expected from the mixing of blackbodies when averaging our
  CMB sky \citep{cs2004}. The negative distortion from adiabatic cooling of
matter is shown in red. $y$ type distortion is clearly dominated by the
contributions, during and after reionization, from the intergalactic medium and
clusters of galaxies, and the early Universe contributions are difficult to constrain.} 
\end{table}

\subsection{Energy released from recombination of plasma}
We should also mention that recombination lines  also create a
  distortion of amplitude $\Delta T/T \sim 10^{-8}-10^{-9}$ \citep[see][for a detailed
  calculation]{rcs2008}. The distorted spectrum would heat the electrons, 
   adding a  $y$-distortion at the time of HeIII$\rightarrow$HeII
  recombination of   $\sim y(6000)\times 10^{-9}\sim
  10^{-12}$. Additionally, the Ly-$\alpha$ and 2s-1s (2-photon decay)
  photons from recombination with energy $\sim 40$eV, $x\sim 30$, escape as they redshift out
  of resonance, (Compton) scatter on electrons and
  heat the  plasma through recoil effect. The heating can be estimated
  using the analytic solution of Kompaneets equation with only the recoil
  effect \citep{arons,is72}. In the limit of small Compton-$y$ parameter
  ($xy\ll 1$),
  the fraction of energy lost by photons at frequency $x$ is $\sim y\times
  x\sim 1/30$, giving an additional $y$-type distortion of $\sim
  (1/4)(40{\rm  eV})\nHe x y/E_{\gamma}\sim 10^{-12}$. {The
    distortions from HeI and HI recombination are much smaller since they
    happen later, when $y$ is much smaller, although the energy released is
    comparable to HeII recombination.}

\section{Conclusions}
Future experiments, such as PIXIE \citep{pixie}, would be able to
constrain/measure 
spectral distortions in the CMB at high accuracy. There are several sources of
spectral distortions possible from  standard and new physics.
Using the results
of experiments like PIXIE to constrain new physics would require precision
calculations of evolution of the CMB spectrum, especially around the blackbody
surface at $z\sim 2\times 10^6$.  So far, precise calculations have only
been possible numerically, although analytic solutions with $5-10\%$ 
precision around the blackbody surface have been available for a long
time.  We have presented new analytic solutions, which take into account
both double Compton scattering (important at high redshifts) and
bremsstrahlung (important at low redshifts).  We also take into account the
non-stationarity of the problem which is important to achieve high precision.
 The new solutions are presented in Eq. \eqref{optold} (ignoring
non-stationarity) and in Eq. \eqref{optnew} (including the non-stationarity of the
problem). Equation \eqref{optnew} gives accuracy of better than $1\%$ in
blackbody visibility at all redshifts. We also present examples from the standard $\Lambda CDM$ model of
cosmology, {which do not require new physics,} illustrating the structure of blackbody surface.   In particular,
electron-positron annihilation and BBN demonstrate the effectiveness of
Compton and double Compton scattering in maintaining equilibrium at high
redshifts. We also point out the  coincidence/degeneracy among  the three
significant sources of distortions in standard cosmology, Bose-Einstein
condensation of CMB, Silk damping and dark matter annihilation. All of these create
distortions which have roughly the same order of magnitude, especially for
low dark matter particle masses. This is remarkable considering that the
three sources of distortions have completely different physical origins.
Bose-Einstein condensation is fixed by standard cosmological parameters,
which are now known with high precision. However, the degeneracy between Silk
damping and dark matter annihilation must be taken into account when using
spectral distortions to constrain the primordial power spectrum or dark
matter parameters.
\bibliographystyle{JHEP}
\bibliography{cmbBB}
\end{document}